# A Module-System Discipline for Model-Driven Software Development

## Sebastian Erdweg[a] and Klaus Ostermann[b]

a   TU Delft, Netherlands

b   University of Tübingen, Germany

**Abstract**   Model-driven development is a pragmatic approach to software development that embraces domain-specific languages (DSLs), where models correspond to DSL programs. A distinguishing feature of model-driven development is that clients of a model can select from an open set of alternative semantics of the model by applying different model transformations. However, in existing model-driven frameworks, dependencies between models, model transformations, and generated code artifacts are either implicit or globally declared in build scripts, which impedes modular reasoning, separate compilation, and programmability in general.

We propose the design of a new module system that incorporates models and model transformations as modules. A programmer can apply transformations in import statements, thus declaring a dependency on generated code artifacts. Our design enables modular reasoning and separate compilation by preventing hidden dependencies, and it supports mixing modeling artifacts with conventional code artifacts as well as higher-order transformations. We have formalized our design and the aforementioned properties and have validated it by an implementation and case studies that show that our module system successfully integrates model-driven development into conventional programming languages.

**Keywords**   module system; software components; dependency tracking; model-driven software development; domain-specific languages

## The Art, Science, and Engineering of Programming

## 1   Introduction

Model-driven development [6, 20] is a pragmatic approach to software development. The central idea behind model-driven development is to describe software systems as a combination of *models* which are syntactic representation of domain knowledge, *metamodels* which describe classes of models, and *transformations* which translate models to models or models to code of some base language. For example, in the model-driven development of a web application, we may find models that describe the processing of user requests as a state machine, models that describe the database schemas of persisted data, models that describe the user interface, models that describe security policies, and so on [28]. It is important to note that, in general, models are just syntactic representations without behavior. Only some selected models have an externally pre-defined semantics (such as state machines). Semantics is given to a model by applying different transformations that eventually produce executable code in some base language or model with pre-defined semantics. Therefore, in model-driven development, it is possible and often useful to apply different semantics to a single model.

However, the idealized description of model-driven development in the previous paragraph is not quite complete. In practice, models, metamodels, and transformations are accompanied by build scripts or configuration files that control the application of model transformations by defining transformation pipelines and selecting appropriate input models. Moreover, there is typically a good share of hand-written code in the base language that acts as glue code for the code generated from models and contains regular algorithmic functionality as well as code for domains that have no dedicated modeling support.

In this paper, we investigate how modeling features (models, metamodels, transformations) can be integrated into a programming language without breaking modularity. Existing model-driven frameworks are not designed in a way that supports modularity, because many module dependencies are implicit. In particular, it is difficult to reason about dependencies regarding generated artifacts. For example, as displayed on the right-hand side, if model A is input to a transformation T producing B, which is used by C, changes to which artifacts can impact C? Clearly, C depends on B and any module imported by B. But since B is a generated artifact (dotted outline), we also have to consider the implicit dependencies (dashed arrows) of B to the generator T and its 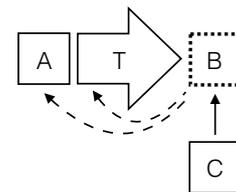 input A. Thus, a change to either A or T can also impact C. To gather this information in an existing model-driven framework, we would have to inspect the build script—and thus the global project structure—to determine which transformation generated B from which input models. Moreover, without further control, transformations can inject arbitrary additional module dependencies into generated code, which prevents static reasoning altogether.

To address these problems, we propose a module system for model-driven development that integrates models, conventional code, metamodels, and transformations as modules into a single framework for managing dependencies. Transformations are applied in import declarations such as **import** x = Trans(Model). This import denotes that the current module requires the result of applying the transformation defined in module Trans to the model defined in module Model. The transformation application results in a generated module, which is locally named x. This makes it easy to mix generated and user-written code without requiring build scripts. Moreover, the explicit declaration of dependencies on generated artifacts via import statements enables our module system to manage the intricate dependencies involved in model-driven development. In contrast to existing model-driven frameworks, we can guarantee that there are no hidden compile-time dependencies, which enables static, modular reasoning and separate compilation.

In analogy with the terminology from model-driven development, we call modules that represent input data for transformations *models*. Inspired by Bezevin's idea that 'everything is a model' [4], every module in our system either *is* a model or can be reified as a model. Since transformations are modules themselves, they can be reified as models, which enables higher-order transformations. We call our proposal, perhaps improvidently, a *module system*, but it is important that this work is only concerned with how dependencies are declared and transformations are integrated into the module system; our work is orthogonal to other aspects of module systems, such as information hiding or static typing.

In summary, we make the following contributions:

- We identify the deficiencies of model-driven development and formulate design requirements from the perspective of modular software development and by reviewing existing user studies (Section 2).
- We propose to organize models and transformations as modules, formalize a corresponding calculus, and prove theorems showing separate compilation and avoidance of hidden dependencies (Sections  3 and 4).
- We present a Java-based implementation of our design that adheres to the formal properties of the calculus (Section 5). The source code of our compiler, IDE, and case studies is available online: http://www.erdweg.org/projects/sugarj/models/.
- We demonstrate that our design and its properties do not impede application to typical model-driven scenarios. We present three case studies that demonstrate that our module system supports common applications of model-driven development, such as imposing new behavior on existing components, externally configurable components, and the composition of transformations (Section 6).
- To the best of our knowledge, this is the first paper to demonstrate that module systems can manage dependencies between generators as well as generated and user-supplied code artifacts as present in model-driven development.

## 2  Problem Statement and Design Goals

This paper addresses problems related to the application of model-driven development. A simple yet typical usage scenario for model-driven development is the modeling of high-level data descriptions and their translation into object-oriented encodings.

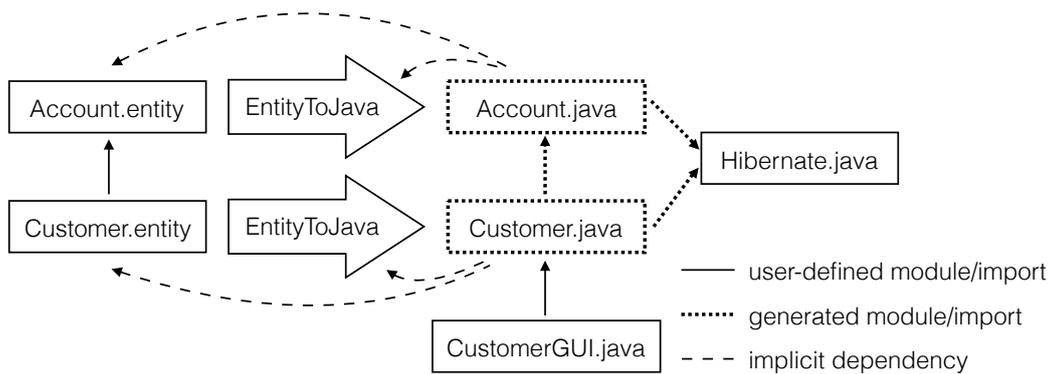

■ **Figure 1**   User-defined, generated, and implicit module dependencies.

Examples of data-description languages include UML class diagrams, EMF metamodels, XML Schema, database schemas, but also many parser generators can produce object-oriented AST encodings from a grammar. We illustrate a simple example in Figure 1.

Figure 1 shows two user-defined entity declarations `Account` and `Customer` on the left, which we assume to be models in some data-description language. The `Customer` model declares a field of type `Account` and therefore has a dependency on the `Account` model. The transformation `EntityToJava` translates each entity declaration into a corresponding Java class, preserving the dependency between `Customer` and `Account` in the generated code. The generated code furthermore imports the user-defined Java library `Hibernate`. Finally, the client program `CustomerGUI` uses the generated code by importing the corresponding Java class.

As shown in Figure 1, this example involves various kinds of dependencies. However, build scripts used in existing transformation frameworks are not aware of all involved dependencies, which leads to the following problems.

**Modular reasoning.**   The client program `CustomerGUI` imports a generated module, which is not present in the original source code. To understand the interface of the imported `Customer` Java class, a programmer ultimately has to read and understand the build script—and thus the global project structure—to identify the original entity declaration and the transformation that produces the Java encoding.

**Separate compilation.**   The generated Java classes `Customer` and `Account` implicitly depend on their original entities and on the generating transformation: A change to an entity or the transformation may entail a change in the generated classes. However, build scripts typically do not model such fine-grained dependencies per module, but instead trigger regeneration for all entities. One could of course alter the build script and hardwire the module dependencies into the build script, but this information replicates information that is already in the source files and may hence become inconsistent. The problem becomes harder yet due to dependencies that are synthesized by the transformation, such as the dependency on `Hibernate` in our example. This dependency is not visible in the original entity models but only in the generated code after executing the transformation, which inhibits static reasoning about module dependencies.

**Non-modular transformation application.**    Build scripts globally specify which transformation to apply to which input model. This is unsatisfactory for intricate dependencies involving transformations, for example, when transformation `EntityToJava` itself depends on user-defined Java classes or the result of a transformation is another transformation. If we want to support such dependencies with a build script, the build script becomes highly complex because it needs to call transformations in the right order on the right input for all involved code-generation tasks.

**Use multiple transformations.**    Client code such as `CustomerGUI` imports generated modules by name. This makes it difficult to use multiple transformations on a single input model, because clients cannot distinguish the generated code by name. For example, assume we have a separate transformation that, given an entity, generates a JSON serializer for that entity. A build script can apply this transformation to `Customer` and `Account` in addition to `EntityToJava`, but no single Java class can use the result of both transformations.

## 2.1  Design Requirements

We want to design a module system that supports models and model transformations without breaking modularity. To derive design requirements for our module system, we generalize above observations, explore existing literature, and rely on the findings of industrial user studies [19, 15].

First, we note that build scripts inject dependencies into generated modules that are hidden from the developer. This violates *communication integrity*, which postulates that each component in the implementation of a system may only depend on those components to which it is explicitly connected. Communication integrity has been recognized as a pillar of modular program understanding and component-based software architecture [27, 23, 1]. Build scripts violate communication integrity because various dependencies remain implicit: The client of a generated component only declares a dependency on the generated artifact, while, actually, the client depends on the original component and the transformation that produces the generated component. In the example above, if either the original component `Customer.entity` or the transformation `EntityToJava` changes, the client `CustomerGUI.java` may be affected because the generated component may have changed. However, these dependencies are hidden inside the build script.

Reasoning about dependencies becomes even harder when considering the dependencies from generated components to other user modules. These dependencies result from transformations that generate import statements. For example, the code generated by `EntityToJava` may contain an import of another module that provides functions for data persistence (e.g., Hibernate). This dependency cannot be seen by considering the dependencies of `EntityToJava` or input entities such as `Customer.entity`. Rather, this dependency is *generated*.

Therefore, the dependency structure is fragile: It depends on implementation details of transformations and their input. Fragile, implicit dependencies are at odds with basic software-architecture principles, in which dependencies are seen as a high-level architectural concern [2]. From this discussion, we derive the first design requirement for our module system: **(R1) The module system must prevent hidden dependencies.**

Hidden dependencies also have negative technical consequences, most notably the loss of separate compilation. That is, if dependencies are not explicit and can arise during the execution of a transformation, build systems must fall back to conservatively rebuilding all potential dependencies in a global fashion.[1] Kuhn et al. [19] report that developers working with model-driven frameworks complain that long build cycles, which often take multiple hours in bigger projects, negatively impact feedback cycles and consequently more effective adoption of code generation. Our second design requirement is: **(R2) The module system must support separate compilation.**

A problem, which has been expressed many times in the literature, is that existing transformation-based approaches to DSLs or model-driven development entail a gap between domain-specific programming and general-purpose programming [24, 15]. The problem is that existing approaches require *three* separate module systems: one for domain-specific programs (e.g., interconnected entity declarations), one for general-purpose programs (e.g., interconnected Java classes), and one for connecting the two (e.g., describe which and how Java classes are generated from entities). This makes it difficult to tightly integrate domain-specific and general-purpose components, and even more so if multiple domain-specific languages are involved. Consequently, it is not surprising that developers are worried that domain-specific and general-purpose code become inconsistent [15]. **(R3) The module system must support tight integration of domain-specific and general-purpose components by a common dependency management.**

A final concern regarding build scripts is the applicability of transformations at different metalevels. Kuhn et al. report that model-driven developers often have no tool support for creating tools similar to the ones they use themselves [19]. Model-driven frameworks such as EMF [29] also suggest a strict stratification into metalevels; for instance, using a generated editor entails starting a new Eclipse instance with a separate workspace. We believe that it is important that everything can be subject to transformation (in the spirit of 'everything is a model' [4]). This includes being able to generate domain-specific programs, general-purpose programs, and transformations themselves, such that the programming model stays the same regardless of the metalevel and higher-order transformations become possible. **(R4) The module system must support transformations uniformly across metalevels.**

Since the above requirements place quite strong restrictions on the design of a module system for model-driven development, it is not clear whether this has a negative impact on the system's expressiveness compared to existing model-driven frame-works. Hence our last requirement is practicability. **(R5) The module system must be applicable for model-driven software development in practical applications.**

---

[1] Build systems with dynamic dependency management can avoid this problem [11], but virtually all build systems in practice use static dependency management.

<div style="display:inline-block;background:#4472c4;color:white;padding:2px 8px;font-weight:bold">3</div> **Informal Overview of the Module System**

To address the above requirements, we propose to unify code, models, metamodels, and transformations as interdependent programming-language modules. In this section, we informally present our design from a user's perspective by using a Scheme-like language (and s-expressions) to express ordinary base code, transformations, and models. In the next section, we formalize the design in a way that abstracts over the choice of base and transformation language. In Section 5, we integrate our module system into Java and add support for custom metamodels, which we already addressed in prior work [12].

**Modules.**  A module is defined with the **module** construct and consists of a list of imports and a list of definitions. For example, here is a module with an empty import list that defines a factorial function:

```
1 (module                                                          factorial
2  (imports)
3  (define fact (lambda (n) (if (= n 0) 1 (* n (fact (- n 1)))))))
```

A module has no name; names are associated with modules externally (e.g., via file names). For clarity, we show the external name of a module in the top-right corner of its definition; the above module is named *factorial*. As this work concentrates on dependency tracking and separate compilation, we do not consider other typical aspects of modules, such as information hiding or typing, and assume that all definitions are exported.

**Models.**  A model is a syntactic reification of a module and can be processed by model transformations. Our module system allows users to reify existing modules such as *factorial* as the s-expression that defines the module. Furthermore, using the **model** construct in a module body, users can define modules that have no behavior but are a purely syntactic representation of some modeling intent. For example, here is a module whose intent is to describe a data schema for account entities.

```
1 (module                                                            account
2  (imports)
3  (model (entity ((IBAN string) (BIC string) (balance number)))))
```

**Model transformations.**  A model transformation is a function that takes one or more models as input and produces a model as output. The flexibility of model-driven development comes from model transformations, because they are defined separately from models and provide semantics to them. Since there can be multiple transformations applicable to a single model, a model can have multiple semantics; the user decides on the semantics by applying an appropriate transformation.

We define transformations using the **transformation** construct. In our example language, a transformation is a regular Scheme function that takes and produces a model in the form of s-expressions. Model transformations receive the complete representation of a module including its imports, and must produce a complete representation of a module including all required imports. For example, we define a transformation that, when applied to module *account* from above, yields a set of functions `make-account`, `account-IBAN`, `account?`, and so forth, which can be used to create and query accounts (and entities in general). We use Scheme-style quasi-quote (') and unquote (,) in the transformation:

```
1 (module                                                    entity-to-record
2  (imports)
3  (transformation (lambda (entity) '(module (imports) ,(trans-body entity))))
4  (define (trans-body entity) ...))
```

**Transformation application.**    A user of our module system can depend on the result of applying a model transformation to a user-defined or reified model. Our system integrates the application of transformations as import statements. Accordingly, the compiler executes a model transformation when compiling the module that contains the import statement. For example, the following module applies transformation *entity-to-record* to the model of *account*, imports the resulting module as `acc-rec`, and can hence use the `acc-rec:` prefix to refer to functions generated by the transformation.

```
1 (module                                                              test
2  (imports (acc-rec (entity-to-record account)))
3  (define checking (acc-rec:make-account "GB29 NWBK 6016 1331 9268 19" "..." 12345)))
```

A single model can be passed to many transformations. For instance, we can define another transformation *entity-to-sql* that transforms entity descriptions into SQL database wrappers. A single module can depend on the application of both transformations to the same entity, using two separate import statements with different local names.

**Higher-order transformations.**    Since we define transformations in modules and every module can be syntactically reified as a model, transformations can transform and produce the definitions of other transformations. Thus, our design naturally supports higher-order transformations. For instance, a transformation can decorate another transformation by, say, adding logging or synchronization to every function generated by the other transformation.

Higher-order transformations are different from higher-order functions in that transformations must accept and produce syntactic representations of modules. Accordingly, higher-order transformations take or produce first-order syntactic values (namely, models) that *represent* transformations.

**Propagating transformations.**    Since modules are interconnected via imports, in a modular design it is inevitable that the transformation of one module entails the transformation of other imported modules. For example, consider a second entity model describing a data schema for customers that have an account:

```
1  (module                                                    customer
2   (imports (acc account))
3   (model (entity ((name string) (address string) (customer-account acc:account)))))
```

According to the schema, each customer has an account. To acknowledge this modeling intent, module *customer* imports module *account*.

Beyond expressing modeling intent, the import also enables that code generated for *customer* can interact with code generated for *account*. For example, function `make-customer` generated by *entity-to-record* should be able to call `acc:account?` to dynamically check that the third argument indeed is of type `acc:account`. Thus, when applying *entity-to-record* to *customer*, we need to propagate the transformation application and adapt the import of *account* to import *(entity-to-record account)* instead, thus making the functions generated for *account* available in the code generated for *customer*. Since a transformation takes and produces comple modules including imports, we can propagate a transformation application by generating further transformation applications in the generated module:

```
1  (module                                              entity-to-records
2   (imports)
3   (transformation (lambda (entity) `(module (imports ,(trans-imports entity)) ,(trans-body
4       ↪ entity))))
5   (define (trans-imports entity) (map trans-import (get-imports entity)))
6   (define (trans-import imp) `(,(car imp) (current-trans ,(cdr imp))))
7   (define (trans-body entity) ...))
```

This transformation generates a module with imports produced by mapping `trans-import` over the original list of imports. Function `trans-import` preserves the local name of the import (`car imp`) but generates a transformation application (`current-trans ,(cdr imp)`). This ensures that all referenced entities are subject to the same model transformation. Our system defines the native module name `current-trans` to refer to the current transformation, because the name of the propagated transformation is externally defined and not available in the body of the transformation itself. Similarly, we define native module names `current-arg-i` for the arguments of the current transformation.

This section gave an overview of our design to organize models and transformations alongside code as modules. To clarify the relation between models, transformations, code, and separate compilation, we present a formal semantics in the subsequent section.

## 4  Semantics of the Module System

After the informal overview in the last section, this section explains the exact semantics of our module system, in particular, how the semantics ensures that no hidden compile-time dependencies can occur and how it enables separate compilation. We specify the semantics of our module system by a compositional mapping of modules to their semantic domain [30]. The semantics of our module system is parametric over the base

```
1   ------------------ syntactic domains -------------------------
2
3   data Module = Module [(String,Import)] Syntax
4   data Import = SimpleImport String | Apply Import [Import]
5   data ModuleType = TBase | TTrans | TModel
6
7   ------------------ semantic domains -------------------------
8
9   data D = DBase Model Base | DTrans Model Trans | DModel Model
10  type ModuleEnv = [(String, D)]
11  type Model = (Module,ModuleEnv)
12  type Trans = [Module] -> Module
13
14  -------------- interface to host language -----------------------
15
16  type Syntax = ...   -- syntactic domain of host language
17  type Base = ...     -- semantic domain of host language
18  compileTrans :: Syntax -> ModuleEnv  -> Trans
19  compileBase :: Syntax -> ModuleEnv -> Base
20  typeOf :: Syntax -> ModuleType
21
22  ------------------ semantic functions ----------------------
23
24  compileModule :: ModuleEnv -> Module -> D
25  compileModule env m@(Module imports body) =
26    let bodyenv = map (\(n,i) -> (n,compileImport env i)) imports
27        model = (m,bodyenv)
28    in case typeOf body of
29         TBase  -> DBase model (compileBase body bodyenv)
30         TTrans -> DTrans model (compileTrans body bodyenv)
31         TModel -> DModel model
32
33  compileImport :: ModuleEnv -> Import -> D
34  compileImport env (SimpleImport n) = fromJust (lookup n env)
35  compileImport env (Apply imp imps) = compileModule genEnv genModule
36    where
37      DTrans ( _, envtrans) dtrans = compileImport env imp
38      tr_args = map (compileImport env) imps
39      arg_models = map getModel tr_args
40      genModule = dtrans (map fst arg_models)
41      argenv = concatMap snd arg_models
42      transargs = zip ["current-arg-"++(show n) | n <- [1..]] tr_args
43      genEnv = transargs ++ [("current-trans",tr)]  ++ argenv ++ envtrans
44
45  getModel :: D -> Model
46  getModel (DBase m _) = m
47  getModel (DTrans m _) = m
48  getModel (DModel m) = m
```

■ **Figure 2**   Semantics of our module system defined by compilation of modules and imports.

```
1   type Syntax = SExp
2   data SExp = St String | Nm Integer | List [SExp] deriving Eq
3
4   type Base = [(Name, Val)]
5
6   compileTrans :: SExp -> ModuleEnv -> Trans
7   compileTrans (List [St "transformation", sexp]) menv =
8       val2module . t . (map module2val)
9       where t = evaltrans ((menv2env menv) ++ initialenv) (parseexp sexp)
10
11  compileBase :: SExp -> ModuleEnv -> Base
12  compileBase (List [St "base", List defs]) menv = res
13    -- the recursive environment allows recursion within a module
14    where
15      res = map f defs
16      env = (res ++ (menv2env menv) ++ initialenv)
17      f (List [St x,d]) = (x,eval env (parseexp d))
18
19  typeOf :: SExp -> ModuleType
20  typeOf (List [St "model", _]) = TModel
21  typeOf (List [St "transformation", _]) = TTrans
22  typeOf (List [St "base", _]) = TBase
23
24  data Exp = Fun [String] Exp | App Exp [Exp] | Quote SExp | ...
25  data Val = Closure ([Val] -> Val) | S String | N Integer | L [Val]
26  type Env = [(String,Val)]
27
28  evaltrans :: Env -> Exp -> ([Val] -> Val)
29  evaltrans env e  = case eval env e of Closure f -> f
30
31  eval :: Env -> Exp -> Val        -- standard evaluator
32  module2val :: Module -> Val      -- reify module
33  val2module :: Val -> Module      -- de-reify module
34  menv2env :: ModuleEnv -> Env  -- reifies all D's in env as values
35  parseexp :: SExp -> Exp          -- parse s-expression into expression
36  initialenv :: Env                       -- standard functions and constants
```

■ **Figure 3**  Example Scheme-like instantiation for base and transformation language.

language and the transformation language. Later in this section, we show an example instantiation of the module system with a Scheme-based base and transformation language.

Figure 2 defines the syntax and semantics of our module system as a set of Haskell definitions. A `Module` has a list of named imports and a body which consists of the `Syntax` of an unspecified base language. An `Import` is either an ordinary import of a module in the module environment, or the application of a transformation to a sequence of other modules. A `ModuleType` stands for one of the three types of modules described in the previous section.

The semantics of each module is described by a value of type `D`, with one alternative for each `ModuleType`. Module environments `ModuleEnv` map names to values of type `D`.

A `Model` is a module closed under the module environment used when the module was compiled. This is similar to how function closures in functional programming close a function definition in the environment in which they were evaluated. Note that every alternative of `D` has a model, which is important to reification of arbitrary modules. A transformation transforms a list of (syntactic) modules to another module.

The module system is parametric over the base and transformation languages. The interface for this parameterization is described in the third block in Figure 2. An example instantiation of this interface for the Scheme-like language we used in the previous section can be found in Figure 3.

The semantics of modules is defined in the two functions `compileModule` and `compileImport`. To compile a module, we first compile (think: resolve) all imports, which yields an environment of modules that may be used in the body. The model of the current module is closed under this environment, so that dependencies within the model are always resolved in the definition context, independent of the client context. Depending on the type of the module, `compileModule` calls either `compileBase`, `compileTrans`, or just returns the model.

Not surprisingly, `compileImport` is the most involved semantic function in the module system, because it has to handle transformation applications. But first, a `SimpleImport` is just looked up in the environment. In the `Apply imp imps` case, we first compile `imp`, which must yield a transformation `dtrans`. Then we compile all arguments to the transformation, yielding `tr_args`, from which we extract the models `arg_models` using `getModel`. We apply the program transformation `dtrans` to the syntactic modules that are stored (as `fst` component of the pair) in `arg_models`, yielding a generated module `genModule`.

The remainder of the code sets up the environment in which the generated module `genModule` must be compiled. The `current-trans`, `current-arg-1`, `current-arg-2`, ... identifiers are bound to the current transformation and its arguments, as described in the "Propagating transformations" part of the previous section. Apart from that, the environment for the compilation of `genModule` consists of the environment of the transformation `envtrans` and the concatenated environments of the transformation arguments `argenv`. In particular, the current environment `env` may not be used during the compilation of `genModule`.

**Example.** Let us revisit the example from Figure 1, which we can compile by the following sequence of calls to `compileModule`:

```
1  compileModule [] (parse "Account.entity") = acc :: DModel
2  compileModule [("Account", acc)] (parse "Customer.entity") = cus :: DModel
3  compileModule [] (parse "Hibernate.java") = hib :: DBase
4  compileModule [("Hibernate", hib)] (parse "EntityToJava.trans") = e2j :: DTrans
5  compileModule [("Customer", cus), ("EntityToJava", e2j)] (parse "CustomerGUI.java") = gui ::
        ↪ DBase
```

First, we compile module `Account.entity` in the empty environment, since it does not depend on any other module. This yields a modeling value `DModel`. Second, we compile `Customer.entity`, which imports `Account.entity`. Since in our design imports within models are significant, the compiler requires a compiled version of the `Account` module in the environment. Third, we compile the base program `Hibernate.java` in the empty environment, assuming it does not depend on any other modules. Fourth, we compile the transformation `EntityToJava`. Since the transformation generates a dependency on `Hibernate`, the transformation must declare a dependency on this module already. Thus, we require `Hibernate` in the environment. Finally, `CustomerGUI.java` depends on the result of applying `EntityToJava` to `Customer.entity`. Even though the code generated by `EntityToJava` has additional dependencies on `Hibernate` and the result of transforming `Account`, these modules are not needed in the environment of `CustomerGUI`. Instead, all dependencies of compiled artifacts are captured in the model closure.

To see how our semantics prevents hidden dependencies, let us assume `EntityToJava` did not import `Hibernate`. Thus, the dependency would appear unexpectedly in the generated module. Our compiler prevents this because, even if we add `Hibernate` to the environment of `CustomerGUI`, compilation fails since the generated module is compiled in the static context of the transformation and input models. Specifically, `genEnv` in function `compileImport` would not bind `Hibernate` and thus the call `compileModule genEnv genModule` would fail.

## 4.1 Metatheory

A key property of our module system is that there can be no hidden dependencies between modules. In particular, a model transformation cannot introduce new dependencies that are not explicitly declared. To state and prove this dependency property, we first define what the declared dependencies of a module are:

```
1  deps :: Module -> [String]
2  deps (Module imps _) = concatMap deps_imp (map snd imps)
3
4  deps_imp :: Import -> [String]
5  deps_imp (SimpleImport s) = [s]
6  deps_imp (Apply imp imps) = deps_imp imp ++ concatMap deps_imp imps
```

The definition collects all module names that are referenced in simple import statements of the module. Our correctness theorem says that the result of compilation depends only on explicitly declared module dependencies.

**Theorem 1 (No hidden dependencies)** For all modules `m` and environments `env` and `env'`, if `lookup x env = lookup x env'` for all `x` in `deps m`, then `compileModule env m = compileModule env' m`.

The proof is straightforward if we can establish the following lemma about `compileImport`.

**Lemma 1** For all imports `i` and environments `env` and `env'`, if `lookup x env = lookup x env'` for all `x` in `deps_imp i`, then `compileImport env i = compileImport env' i`.

**Proof 1** The proof is by induction on the structure of `i`. There are two cases. 1) If `i` is `SimpleImport x`, then `x` is in `deps_import i` by the definition of `deps_import`, and thus `lookup x env = lookup x env'`. 2) If `i` is `Apply imp imps`, then, by induction hypothesis,

envtrans = envtrans', dtrans = dtrans', and tr_args = tr_args'.[2] Since all used functions are pure (depend only on their input) and the remainder of the definition does not depend on `env`, we can reason that `genModule = genModule'` and `genEnv = genEnv'`. Specifically, the environment in which we compile the generated module, `genEnv`, does not depend on `env` beyond the explicitly declared dependencies. Since the inputs to `compileModule` are the same in both cases, so must be its result.

As direct consequence of the avoidance of hidden dependencies, we obtain separate compilation (R2): A module can be compiled in separation of modules that it does not directly depend on.

**Theorem 2 (Separate compilation)**  For all modules `m`, environments `env`, and module names `n`, if `n` not in `deps m`, then `compileModule env m = compileModule (delete n env) m`.

## 5    Models and Transformations as Modules for Java

We realized models and transformations as modules in the Java-based programming language SugarJ. Like Java, SugarJ is a compiled language and modules play a double role as compilation units and types. In this section, we report on how to integrate models and transformations as modules into SugarJ.

**Background on SugarJ.**  SugarJ [12, 8] is a syntactically extensible variant of Java that allows programmers to define new syntactic constructs. A SugarJ extension consists of (i) context-free productions that extend the grammar of the base language and (ii) a desugaring transformation that translates the new syntactic constructs into existing constructs of the base language. In addition to syntax and desugaring, a SugarJ extension can define static checks and editor services for new syntactic constructs [10].

SugarJ organizes language extensions as base-language modules that contain the corresponding productions and desugaring rules. A developer can activate a SugarJ extension using a regular import statement. When the SugarJ compiler encounters such an import of a language extension, it adapts the parser and desugaring for the remainder of the module. In contrast to the module system proposed in this paper, SugarJ did not support models or model transformations and desugarings were preemptively executed by the compiler. However, we reuse SugarJ's support for grammar extensions for the definition of metamodels featuring concrete modeling syntax.

**Adding models to SugarJ.**  We augmented SugarJ with support for models and model transformations as modules. First, we added a construct for users to define models using *abstract syntax*. For example, we can define entities like in Section 3 in SugarJ as follows:

---

[2] We use the convention that envtrans', dtrans', and so forth, denote the values of the local variables in `compileImport` for the `env'` case.

```
1  package banking;                                    banking.Account
2  public model Account instantiates Entity {
3    [Property(["IBAN", "String"]), Property(["balance", "Integer"]), Property(["pin", "String"])] }
```

However, due to SugarJ's syntactic extensibility, developers can define a metamodel
with *concrete syntax* for models. For example, we provide the following concrete syntax
for entities:

```
1  package banking;                                    banking.Account
2  import entity.Metamodel;
3  public entity Account {
4    IBAN :: String,   balance :: Integer,   pin :: String }
```

The library `entity.Metamodel` defines the concrete syntax and a desugaring to the
abstract syntax above. Concrete syntax is immediately desugared when compiling
a model and does not affect the rest of our design. To reify modules as models, we
changed the SugarJ compiler to store a module's desugared abstract syntax alongside
other compiler outputs (such as *.class* files).

**Adding transformations to SugarJ.**   SugarJ employs Stratego [31] for desugarings, and
we reuse Stratego as a model transformation language. For example, the following
code translates an entity into a Java class:

```
1  package entity;                                     entity.ToRecord
2  public transformation ToRecord {
3    main = CompilationUnit(map(f))
4
5    f = if ?ModelDec(ModelDecHead(_, _, Id("entity.Metamodel"), _))
6      then entity-to-java-record
7      else if ?TypeImportDec(type)
8        then !TransImportDec(None,
9            TransApp(Id(<current-transformation-name>), type))
10       else id end end
11
12   entity-to-java-record = ... }
```

A transformation must define a `main` strategy. The above transformation expects
a compilation unit as input and maps function `f` over the contained declarations.
Function `f` calls `entity-to-java-record` in case the declaration is an entity model. If instead
the declaration is an import statement, `f` generates a transformation application of
*entity.ToRecord* to the imported module. The actual transformation is slightly more
involved due to error handling.

Like in the formalization, users can apply transformations via import statements.

```
1  import AccountRecord = entity.ToRecord(entity.Account);
```

The local name (here `AccountRecord`) is optional; by default the model name (here
`Account`) is used. In contrast to the formalization, our implementation in SugarJ also
supports circular module dependencies. In particular, models and generated modules
can depend on each other circularly. The compiler detects and resolves mutually
dependent modules by handling them simultaneously.

**Rejecting hidden dependencies.** One considerable difference between the formalization and the implementation of models and transformations in SugarJ is in the management of dependencies. Our formalization has an explicit environment of resolvable modules and rejects hidden dependencies because they cannot be resolved. In contrast, our SugarJ integration follows an approach more traditional to compilers, namely to look up required source files on the hard drive and to compile them on demand. Thus, dually to module environments, our SugarJ implementation detectes hidden dependencies based on *compilation summaries* that contain the following information:

- A field `used` that stores the names of modules used during compilation.
- A field `genBy` that stores the names of modules that were involved in the generation of the current module. For non-generated modules, `genBy` is the empty set.

After the SugarJ compiler finishes processing of a generated module, it checks the compilation summary to identify and reject hidden dependencies. We compute the hidden dependencies `hidden(mod)` of a generated module `mod` as follows:

```
1  allowed(mod) = mod.genBy ∪ {m.used | m ∈ mod.genBy}
2  depOK(m, mod) = m ∈ allowed(mod) ∨ (m.genBy ≠ ∅ ∧ ∀ m'∈m.genBy. depOK(m', mod))
3  hidden(mod) = {m | m ∈ mod.used, ¬ depOK(m, mod)}
```

The set of `allowed` module dependencies corresponds to the environment used for compiling generated modules in the formalization. A module may refer to the transformation, its inputs, and their direct dependencies. A dependency from a module `mod` to a module `m` is OK if `m` is one of the allowed modules or if `m` is generated and all modules `m'` involved in the generation are OK dependencies of `m`. The latter alternative means that a generated module may refer to other generated modules if all modules required for the generation of these modules may be referenced. The set of `hidden` dependencies then is the subset of `used` that are not OK. The SugarJ compiler indicates an error if there is any hidden dependency, that is, if `hidden(mod)` is not empty.

**Modules as types.** Java modules also serve as types. For example, a class declaration defines a class type, which can occur in code as a type annotation. Since we use model transformation to generate Java modules, the question arises how this influences typing. In particular, when we apply different transformations to the same model, do the generated modules have the same Java type?

For SugarJ, we decided that the type of a generated module should be uniquely determined by the applied transformation and its input models. Thus, for example, all references to the generated module `entity.ToRecord(Account)` correspond to the same Java type. However generated modules `entity.ToRecord(Account)` and `entity.ToSql(Account)` have different types (they may implement a common interface). To encode this type-equality relation on top of Java, the SugarJ compiler automatically renames generated modules and references to them.

```
1  public mapping EventMapping {                          orm.persistor.test.EventMapping
2    table "EVENTS";
3    private int nextId = 0;  // a mapping can have local state
4    column EVENT_DATE: long = {
5      persist(e) = e.getDate().getTime();
6      load(e, stamp) = e.setDate(new Date(stamp)); }
7    column TITLE: String = {
8      persist(e) = e.getTitle();
9      load(e, t) = e.setTitle(t); }
10   synthesized column ID: long = {
11     primary key;
12     init = { nextId = nextId + 1; nextId }; } }
```

■ **Figure 4**  Separate specification of an ORM using our DSL.

## 6 Case Studies

We designed a module system to manage models and transformations while avoiding hidden dependencies and providing separate compilation and uniformity. In Section 4, we have formally validated that our module system avoids hidden dependencies (R1) and achieves separate compilation (R2). In this section, we experimentally validate support for integration of code and models (R3), uniformity (R4), and practical applicability (R5). Overall, we demonstrate that pragmatic use cases of model-driven development are compatible with desirable module-system features.

### 6.1 Imposing new Behavior on Existing Components

One strength of model-driven development is the possibility to produce multiple semantic artifacts from the same syntactic artifact. Our module system inherits this strength: We can reify the model of any module and we can use transformations to describe new alternative semantics for these modules.

To demonstrate this feature, we have designed and implemented a DSL for declaring object-relational mappings (ORM). An ORM specification desugars to a model transformation that we can superimpose on existing Java classes. For example, Figure 4 shows a simple ORM specification that derives two columns from fields of the original class and synthesizes a third column. When applying an ORM specification to a Java class, the transformation produces a new Java class that extends the original Java class with (i) synthesized fields, (ii) conversion functions, and (iii) a pair of specialized accessor functions for each targeted column. Due to inheritance, the produced class can transparently be used in place of the original class.

The mapping contains essentially the same information that user's of the Java Persistency API specify using annotations, but our mapping defines conversion functions for dates and the generation of IDs programmatically rather than by convention. Moreover, our DSL separates persistency concerns from class declarations. While these concerns are defined in separate files (similar to Hibernate), their dependencies are managed

```
1  package graph;                    graph.GraphFeatureModel
2  import variability.model.Syntax;
3  public featuremodel GraphFeatureModel {
4    features EdgeImpl,OnlyNeighbors,NoEdges,...
5    constraint Edges && EdgeImpl
6    constraint Edges -> (Directed xor Undirected)
7    constraint EdgeImpl -> (OnlyNeighbors xor NoEdges)
8  }
```

**(a)** Feature model: features and constraints.

```
1  package graph;                    graph.GraphFeatureModel
2  import variability.config.Syntax;
3  import graph.GraphFeatureModel;
4  import variability.CheckConfig(graph.GraphFeatureModel);
5  public config DirectedNeighbors for GraphFeatureModel {
6    enable EdgeImpl, OnlyNeighbors, Directed, Weighted, ...
7    disable NoEdges, Undirected, ... }
```

**(b)** Feature config: selects features and satisfies constraints.

```
1   #ifdef(NoEdges)                                          impl.Graph
2   public EdgeIfc addEdge(Vertex start, Vertex end, #ifdef(Weighted) int weight) {
3     start.addAdjacent(end);
4     #ifdef(Undirected) end.addAdjacent(start);
5     #ifdef(Weighted) start.setWeight(weight);
6     #ifdef(Undirected && Weighted) end.addWeight(weight);
7     return (EdgeIfc) start; }
8   #ifdef(OnlyNeighbors)
9   public EdgeIfc addEdge(Vertex start, Vertex end, #ifdef(Weighted) int weight) {
10    Neighbor e = new Neighbor(end, #ifdef(Weighted) weight);
11    addEdge(start, e);
12    return e; }
```

**(c)** A configurable Java class as a model. In Java, the methods would be duplicate and yield a compile error.

■ **Figure 5**   Our system enables separate compilation of externally configurable components.

by the same module system (unlike Hibernate). This also makes it possible to apply alternative database mappings to a class and use different mappings side-by-side.

This case study shows how model-driven techniques can add functionality to existing components, thus separating concerns. Our module system makes it easy to apply mappings locally, keeping track of the actual dependencies. In fact, our mapping DSL is fully interoperable with other frameworks implementing the Java Persistency API: Manually annotated classes and mapped classes can be persisted in the same database and the mapping of a manually annotated class can be factored out without invalidating entities in the database.

### 6.2 Externally Configurable Components

As a second case study, we have built a framework for `#ifdef`-based feature-oriented software development (FOSD) [3] using models and transformations in SugarJ. A FOSD product line consists of (i) a feature model that declares available features (named configuration options) and constrains their combination, (ii) feature configurations that determine the activated features and that adhere to the constraints of the feature model, and (iii) configurable components that express conditionally included code fragments using `#ifdef` statements over the names of declared features.

FOSD with `#ifdef` is an interesting case study because it is widely used in practice and because the deep integration of normal program code, `#ifdef` conditionals, and feature configurations is challenging from a language integration point of view. We encode a configurable Java component as a model that connects to other configurable and non-configurable Java modules through import statements. A feature configuration then corresponds to a transformation that transforms a configurable component into a regular Java module by selecting the relevant branches of an `#ifdef`. Our case study goes beyond traditional `#ifdef`-based variability management. With a technology such as the CPP preprocessor, the whole project is a single big product line. In contrast, our case study supports local configuration of configurable components since configurations are applied in import statements that do not have a global effect. This enables developers to use variable components multiple times with different configurations within a single software project. Moreover, our case study supports partial configuration where only some `#ifdef`'s are eliminated.

To demonstrate our encoding of FOSD, we implemented a configurable graph library proposed by others as a standard benchmark for FOSD [21]. Figure 5 shows excerpts of the library's feature model, a configuration, and part of a class of the product line. The feature model declares that every variant must support the `Edges` and `EdgeImpl` features, which entail exactly one of `Directed` or `Undirected`, and so on. The feature configuration `DirectedNeighbors` (Figure 5b) selects and deselects features and satisfies the constraints of the feature model, as checked by the generated static analysis `variability.CheckConfig(GraphFeatureModel)` using a SAT solver.

As this case study illustrates, our design is well-suited to encode externally configurable software for four reasons. First, our design supports separate compilation and separate checking of modules: Each configurable component can be analyzed and compiled given only its direct dependencies. In particular, we can analyze models prior to configuration to Java code. In our case study, we check that only declared features are used in `#ifdef`'s, but in principle we could, for example, employ a variability-aware type system to guarantee well-typing of configured components [17]. Second, using imports, configurable components (models) can be easily integrated into non-configurable components (code) and vice versa. Third, our design supports local configuration of external components through import statements, which allows different configurations of a component to be used simultaneously. Fourth, the uniformity of our design enables domain-specific abstractions for feature models and feature configurations, which we translate to static analyses and model transformations, respectively.

### 6.3 Composing Model-Driven Language Abstractions

Model-driven development allows developers to define customary modeling DSLs, which is one of the main motivations for the application of model-driven development. For building large sophisticated DSLs, it is important to support the composition of metamodels and transformations of smaller DSLs. We demonstrate that our design supports DSL composition by composing a simple state-machine DSL with the entity language mentioned in Sections 3 and 5. The composed DSL supports state machines with guarded transitions that depend on and manipulate entities.

Like the entity language, the metamodel and transformation of the simple state-machine DSL are realized as modules and the DSL is independently usable. Figure 6a shows an example state machine, which can be used as follows:

```
1  import statemachine.Simulator(banking.ATM);
2  import IO = statemachine.StatemachineIO(banking.ATM);
```

```
1  package banking;                                                        banking.ATM
2  import statemachine.Metamodel;
3  public statemachine ATM {
4    initial state Init
5    events InsertCard, Cancel, PinOK, PinNOK, MoneyTaken, CardTaken
6    state Init { InsertCard => Withdraw, Cancel => Init }
7    state Withdraw { PinOK => PayOut, PinNOK => Revoke, Cancel => Init } ...}
```

**(a)** A simple state machine describing an ATM.

```
1   package banking;                                                   banking.DataATM
2   import statemachine.data.Metamodel;
3   import banking.Account;
4   public statemachine DataATM {
5     initial state Init
6     data acc :: Account, tries :: Integer
7     events InsertCard(Account), Cancel, Pin(String), ...
8     state Init {
9       InsertCard(card) => Withdraw { acc := card; tries := 0 }}
10    state Withdraw {
11      Pin(p) if p == acc.pin  =>  HowMuch,
12      Pin(p) if p != acc.pin && tries < 2  =>  Withdraw { tries := tries + 1 },
13      Pin(p) if p != acc.pin && tries >= 2  =>  RevokeCard,
14      Cancel => ReturnCard } ...}
```

**(b)** A data-dependent state machine describing an ATM.

■ **Figure 6**  State machine DSLs in SugarJ

```
3  ...
4  ATM machine = new ATM();
5  machine.step(IO.parseEvent(machine, "InsertCard"));
6  machine.step(IO.parseEvent(machine, "Cancel"));
7  assert (machine.currentState() == IO.parseState(machine, "Init"));
```

Here, transformation statemachine.Simulator generates a Java class with two methods step and currentState, whereas transformation statemachine.StatemachineIO generates a Java class that provides auxiliary methods for parsing and printing the names of events and states.

We want to compose this state-machine DSL with the entity language to build a DSL of data-dependent state machines that can manage and act upon internal as well as event-supplied data. For example, DataATM in Figure 6b declares internal data fields using the data keyword. The acc field stores the account that is served during a withdrawal, while field tries totals the number of failed pin requests. The transition functions can access this internal data as well as the event-supplied data in preconditions following the if keyword. A transition can mutate internal data as a side effect, which is specified in curly braces. For example, the Withdraw state compares the supplied pin p to the one stored in the account entity and either accepts it, rejects it and increases tries, or rejects it and revokes the card.

Let us discuss how our system supports the composition of models and model transformations. SugarJ also supports the composition of metamodels, as discussed elsewhere in detail [9]. The model DataATM from Figure 6b reuses the account entity banking.Account via an import. Essentially, this import links the two models together, allowing a model transformation to consider the model and generate code for Account when transforming DataATM. In particular, without the import statement, a transformation applied to DataATM would not be allowed to refer to Account because this would be a hidden dependency. This is why model composition through imports plays an important role in our system.

We want to reuse model transformations statemachine.Simulator and entity.ToRecord in building a model transformation statemachine.data.Simulator for data-dependent statemachines, thus enabling the following example code:

```
1  import entity.ToRecord(banking.Account);
2  import statemachine.data.Simulator(entity.ToRecord)(banking.DataATM);
3  ...
4  DataATM machine = new DataATM();
5  Account acc = new Account(1, 30000, "xxxx"); machine.step(machine.event_InsertCard(acc));
6  for (int i=0; i<3; i++) machine.step(machine.event_Pin("WRONG"));
7  assert (machine.currentState()==machine.state_RevokeCard() &&
8          account.getBalance()==30000);
```

Transformation statemachine.data.Simulator takes a transformation for entities as the first argument, which it uses to compile any occurrences of entities within the data-dependent state machine it is applied to. For example, the import of banking.Account within DataATM becomes entity.ToRecord(banking.Account). Technically, the higher-order transformation statemachine.data.Simulator transforms data-dependent state machines to Java by running an augmented version of statemachine.Simulator first and then weaving the new features into the resulting Java code.

This case study highlights two important aspects. First, the composability of transformation directly depends on the uniformity of our design, which makes it possible for a transformation to transform other transformations. This way we can parameterize the semantics of data-dependent state machines over the semantics of entities. Second, since models are modules, it is natural to declare dependencies between models explicitly using imports (e.g., the dependency from DataATM to Account). In fact, our module system not only enables such dependency declarations but even enforces them, because generated code that refers to Account would be rejected otherwise due to a hidden dependency. This is the cornerstone for modular reasoning and separate compilation.

### 6.4 Summary

We designed our module system to enable principled dependency management between source-code artifacts in the face of models and model transformations. Specifically, our module system provides fundamental modularity features (avoid hidden dependencies, separate compilation, modular reasoning) in combination with the flexibility of model-driven development (model transformation, integration of code at different abstraction levels, code generation across all metalevels).

By design, our module system satisfies the component-integration requirement (R3) and the uniformity requirement (R4) from Section 2. To validate that our module system satisfies communication integrity (R1) and separate compilation (R2), we formalized its semantics and verified corresponding theorems. Despite these strong requirements, our module system is expressive and applicable to a large range of problems (R5) as our case studies indicate.

**Typed transformations.** Our module system integrates the application of model transformations into a programmer's modules via import statements. With model transformations being an integral part of programs, an important question surfaces: What is the type of a model transformation, such that clients can be checked without executing the transformation?

For example, for the entity language, transformation `ToRecord` generates getter and setter functions for each property. Accordingly, the type of the generated code directly depends on the input model. To address this in future work, we could try to use a second transformation that generates type information from the input model, and use this type-generating transformation as the type of the code-generating transformation. For example, we can write a transformation that generates Java interfaces from entities to match the type of generated Java classes. We can do the same for configurable Java classes shown in Section 6.2. The problem with such design would be that type-generating transformations resemble dependent types that are asymmetric (mapping from terms to types). For example, for a transformation application $f(g(m))$, if the type transformation $f_T$ of $f$ requires a model as input, we would have to execute the model (not type) transformation $g(m)$, which contradicts our requirements of uniformity.

To satisfy uniformity, instead of taking a model as input, the type of a transformation must take the type of a model as input and produce the type of the generated code. Thus, the type transformation $g_T$ maps the type of $m$ to another type, which is the input of $f_T$. But, traditionally, the type of a model is simply the metamodel it adheres to, which only constrains the syntactic structure of the model. However, to check clients we need *semantic* information about generated code, such as the existence of certain methods. In future work, we will explore a type system that bridges syntactic and semantic types.

## 7 Related Work

Our design is related to earlier work on build scripts, module systems, and metaprogramming with macros and models.

**Module systems and metaprogramming.** Our design focuses on dependency management in the context of model-driven development. This is why it lacks many other typical features of module systems, such as explicit import/export interfaces. Many existing module systems avoid hidden dependencies and feature separate compilation like we do. However, these properties are significantly harder to achieve in the context of code generation, where the dependencies that arise during code generation must be taken into account. All existing module systems we are aware of do not support code generation. In particular, functors as supported by SML [25] do not generate code but are parameterized modules whose meaning is defined abstractly in terms of another module's meaning. In contrast, our transformations are based on inspection and generation of the *syntactic* structure of a module.

Like most module systems, our system distinguishes the module language (module declarations, import statements) from the programming language that constitutes the body of modules. In contrast, programming languages with first-class modules (such as Newspeak [5]) support module declarations as part of regular application code. Paired with code-introspection mechanisms, it might be possible to represent transformations as modules in such languages, too. However, it is not obvious how to ensure there are no hidden dependencies and how to provide separate compilation in this case.

There is a lot of work (such as [22, 16, 18]) on making program transformations safe (e.g., check once and for all that only well-typed code is generated). It would be useful to incorporate those approaches into ours in order to detect transformations that inject hidden dependencies without executing them. Moreover, we want to explore whether previous work on safe program transformations can be used to provide interfaces for transformations and how this fits into our module system.

**Macros and models.**   Macro systems such as the one in Racket were an important inspiration for this work, because they illustrate how code transformation can be tightly integrated into programming languages, including explicit dependencies [13], even between submodules [14]. The main difference between macros and this work is that each macro argument is coupled to a specific macro invocation; there is no notion of a syntactic model whose existence is independent of any particular compile-time transformation (our requirement (R5)). Moreover, existing macro systems do not prevent hidden dependencies, for example, from macros that expand to import statements.

In comparison with existing model-driven frameworks, our design is an attempt to fully bridge modeling and programming, since models and normal program code are organized with the same module system. As support for this statement, we compare SugarJ to Xtext from the Eclipse Modeling Project and the Meta Programming System (MPS) as representatives.

Xtext [7] supports textual domain-specific notation and a template-based transformation language. The application of transformations is specified in an application-specific build script called workflow. A workflow is a global, sequential description of which metamodel to use for parsing a model, which transformation to apply to which model, and how models are connected to (possibly generated) artifacts. As a consequence, Xtext is not modular: Dependencies are not explicit in a module and the framework does not support separate compilation of models. Furthermore, Xtext does not provide a uniformly applicable modeling mechanism: It cannot be used to provide an alternative transformation language.

The Meta Programming System (MPS) [32] avoids parsing and uses projectional editing to modify models directly. In MPS, dependencies between modeling artifacts are specified within a property dialog for each artifact separately. However, these dependencies are not part of the textual projection of an artifact. Furthermore, the application of a model transformation is not specified as part of the client code, but within the original model. Accordingly, when a new client requires a different transformation, this has to be specified in the property dialog of the original model. Finally, MPS does not automatically deduct the set and order of modules for recompilation.

**Build scripts.** Build scripts are similar to our design because they also combine code generation with dependency management. Many build-script technologies exist and are used in practice, for example, GNU Make, Apache Ant, Shake [26], pluto [11], or workflows in model-driven frameworks. However, as discussed in Section 2, build scripts implement dependencies that are external to the module definitions a programmer reasons about. In contrast, our design incorporates program transformations as modules and uses import statements as its single dependency mechanism. This way, our design is able to prevent hidden dependencies and to support separate compilation without the danger of inconsistent dependency declarations in build scripts.

## 8    Conclusion

Model-driven software development is a pragmatic approach to software development that is widely applied in practice yet lacks important modularity features such as separate compilation. In this paper, we presented the design of a module system that incorporates models and model transformations as modules and the application of transformations within import statements. This enables precise dependency management despite dependencies from and to generated artifacts. In particular, our module system guarantees the absence of hidden dependencies. Our case studies demonstrate that the properties of our module system do not impede applicability in model-driven scenarios.

**Acknowledgements**   We would like to thank Michael Eichberg, Christian Kästner, Tillmann Rendel, Guido Salvaneschi, Gabriele Täntzer, Thomas Thüm, Markus Völter, and the reviewers for valuable feedback on earlier versions of this work, and Jonas Pusch for technical assistance with the graph product line.

## About the authors

**Sebastian Erdweg** Sebastian is an assistant professor at TU Delft, where he works on the foundation and application of programming languages. Sebastian received degrees in computer science from TU Darmstadt (BSc 2007), Aarhus University (MSc 2009), and Philipps-Universität Marburg (PhD 2013) and has worked as a postdoctoral researcher at TU Darmstadt until 2016. Please see his website (http://erdweg.org/) for further information.

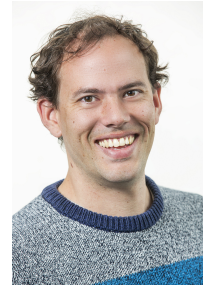

**Klaus Ostermann** Klaus Ostermann is a full professor at the University of Tübingen. His work focuses on programming techniques, tools, and languages. Please see his website (http://ps.informatik.uni-tuebingen.de/team/ostermann/) for further information.

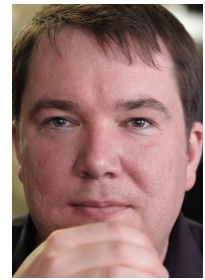